\documentclass[preprint,11pt]{revtex4-1}
\usepackage[utf8]{inputenc}
\usepackage[T1]{fontenc}
\usepackage{bm}
\usepackage{amsmath,ulem}
\usepackage{amsfonts}
\usepackage{amssymb}
\usepackage{graphicx}
\usepackage{siunitx}
\usepackage{float}
\makeatletter
\let\saved@includegraphics\includegraphics
\AtBeginDocument{\let\includegraphics\saved@includegraphics}
\makeatother
\RequirePackage{siunitx}
\usepackage{xcolor}

\definecolor{myBlue}{rgb}{0.1,0.1,0.7}

\newcommand{\Dunit}{\si{\micro\metre\squared \, \second^{-1}}}
\newcommand{\Drunit}{\si{\radian\squared \, \second^{-1}}}
\newcommand{\Vunit}{\si{\micro\metre \,\second^{-1}}}
\newcommand{\Iunit}{\si{\micro\watt \, \micro\metre^{-2}}}
\newcommand{\Muunit}{\si{\micro\metre\squared\,({\second\,\kelvin})^{-1}}}

\begin{document}
\title{Regulated polarization of active particles in local osmotic flow fields}

\author{Lisa Rohde}
    \affiliation{Molecular Nanophotonics Group, Peter Debye Institute for Soft Matter Physics, Leipzig University, 04103 Leipzig, Germany}
\author{Desmond J. Quinn}
    \affiliation{Molecular Nanophotonics Group, Peter Debye Institute for Soft Matter Physics, Leipzig University, 04103 Leipzig, Germany}
\author{Diptabrata Paul}
    \affiliation{Molecular Nanophotonics Group, Peter Debye Institute for Soft Matter Physics, Leipzig University, 04103 Leipzig, Germany}
\author{Frank Cichos}
    \email{cichos@physik.uni-leipzig.de}
    \affiliation{Molecular Nanophotonics Group, Peter Debye Institute for Soft Matter Physics, Leipzig University, 04103 Leipzig, Germany}

\begin{abstract}
Regulation to a well-defined target state is a fundamental requirement for achieving reliable functionality in living systems and maintaining specific non-equilibrium states. The control of certain properties and functionalities of systems on the microscale presents a particular challenge since thermal fluctuations and environmental perturbations dominate. While synthetic active matter has demonstrated remarkable self-organization capabilities, examples of autonomous regulation processes at the single-particle level remain scarce. \\
Here, we show that the interplay of two non-equilibrium processes leads to a regulated polarization state of active particles in local osmotic flow fields. The balance between thermophoretic repulsion and attraction by thermo-osmotic boundary flows, both generated by a single heat source, yields a steady state at which active particles encircle the heat source at a distance that depends on the temperature of the heat source. The balance of both temperature-induced processes causes a polarization of the active particles that is independent of the heat source temperature. The individual control of heat source and active particles in the experiment allows a detailed investigation of the self-regulated polarization effect in which we find that hydrodynamic interactions dominate. As the effects rely on osmotic flows and phoretic interactions, we expect that the observed phenomena can be generalized to other active systems and flow fields.
\end{abstract}

\maketitle

\section*{Introduction}
Robustness in natural and engineered systems is often achieved by sophisticated regulatory mechanisms to maintain specific non-equilibrium states under varying conditions \cite{kitano2004biological, chen2020expanding, von2000segment, csete2002reverse}. From cellular homeostasis \cite{GOMES2015110, hsu2002regulation} to industrial process control, regulation emerges as a fundamental principle for achieving reliable functionality. In microscale systems, where thermal fluctuations and environmental perturbations dominate, achieving robust control of particle properties presents particular challenges. Although synthetic active matter systems have shown significant potential for self-organizing abilities, examples of autonomous regulation processes at the single-particle level are limited.

A future promising approach to study simple regulatory mechanisms can involve the use of artificial active colloids -- synthetic particles that continuously consume energy to maintain their non-equilibrium state.
These self-propelled particles generate motion through asymmetric distributions of chemicals, reactants, or temperature on their surfaces \cite{sanchez2015chemically, bechinger2016active, jiang2010active}. As model systems, these microswimmers provide insights into fundamental questions of self-organization and assembly in active matter \cite{trivedi2022self,araujo2023steering,khadka2018active,palacci2013living,cates2015motility}. Recent advances have demonstrated the construction of complex architectures through controlled interactions with light fields and the exploitation of phoretic and osmotic effects \cite{aubret2018targeted,palacci2013living}. While significant progress has been made in programming collective behaviors like clustering and phase separation through inter-particle interactions \cite{aubret2021metamachines, grossmann2020particle, evans2011orientational,moradi2021spontaneous}, the development of regulatory mechanisms responding to environmental cues remains a key challenge. The ability to control active particles precisely together with external cues provides an indispensable tool for studying the emergence of regulatory processes in active matter systems.

In many biological processes such environmental cues, e.g., composition gradients lead to a steering of self-organization or growth processes. Chemical gradients, for example, can act as directional signals for the organization of microtubules to initiate cell movement \cite{devreotes2003eukaryotic} and are responsible for the assembly of the mitotic spindle around the chromosomes \cite{karsenti2001mitotic,bastiaens2006gradients,prosser2017mitotic}. They are also responsible for maintaining the polarization of cell movement or growth \cite{jain2020role,nabi1999polarization,verkhovsky1999self}.

Although synthetic active particles lack the sophisticated internal chemical networks found in living systems, they exhibit responsive behaviors to various environmental stimuli as well. For instance, active Janus particles demonstrate chemotactic-like responses to chemical gradients analogous to bacterial navigation through tumble rate modulation \cite{popescu2018chemotaxis,vinze2021motion,liebchen2018synthetic,alexandre2004ecological,friedrich2007chemotaxis}. In a similar vein, self-thermophoretic active particles display thermotactic behavior in temperature fields \cite{thermotaxisBregulla}. The asymmetric hydrodynamic flow fields generated during their propulsion \cite{bickel2013flow,campbell2019experimental} enable these synthetic particles to respond to external flows \cite{katuri2018cross, dey2022oscillatory, costanzo2012transport, stark2018artificial} and adapt their dynamics near boundaries where hydrodynamic conditions are modified \cite{zottl2014hydrodynamics,schaar2015detention, fily2014dynamics,shen2018hydrodynamic, uspal2015self}. Here we demonstrate a regulatory mechanism in synthetic active matter where self-thermophoretic Janus particles undergo orientational control through their interaction with localized temperature and thermo-osmotic flow fields. Using a light-absorbing silica particle as a tunable heat source, we generate controlled temperature gradients that create well-defined thermo-osmotic boundary flow fields. The active particles spontaneously organize into stable orbital configurations around the heat source, maintained through a balance of thermophoretic and hydrodynamic interactions. Through independent optical control of both the Janus particle activity and the local temperature field, we reveal how the interplay between particle motility and environmental cues leads to emergent regulatory behavior. The stable orbit maintains the active particle at a position of constant shear flow, which stabilizes the polarization of the spherical Janus particles. The heat source temperature systematically controls the equilibrium orbiting distance, while remarkably, the active particle polarization strength remains invariant. This regulatory mechanism, contrary to previously reported thermotaxis behavior, demonstrates how basic physical interactions can produce robust self-organized states in active matter. Given that the underlying mechanism relies on general phoretic and osmotic effects, we anticipate that similar non-equilibrium self-organization and regulation principles could extend to other particle systems and environmental gradients.

\begin{figure*}[ht]
    \centering
    \includegraphics[width=0.8\textwidth]{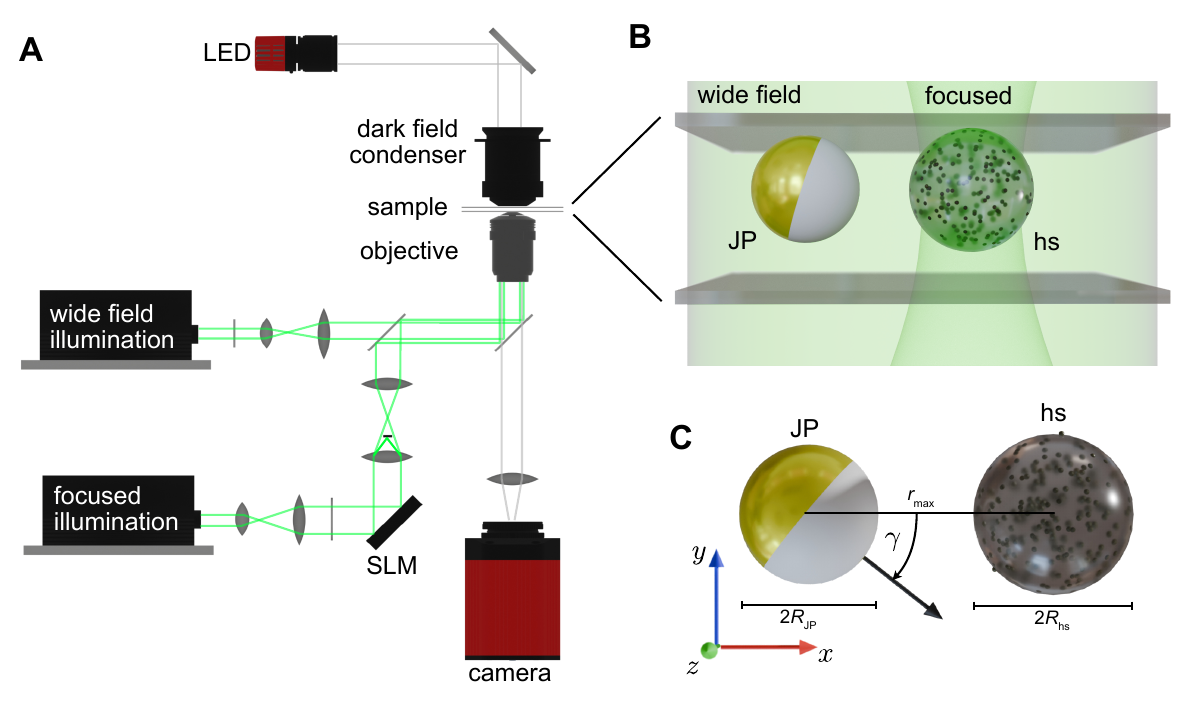}
    \caption{\textbf{Experimental Realization} \textbf{A} Sketch of the experimental setup shows how two laser paths are built in and coupled together. Both lasers can be operated independent from each other. \textbf{B} The iron oxide particle is heated with a focused laser beam. The Janus particles are self-propelled by illuminating the whole field of view with a second laser. The particles translation is restricted to a two dimensional motion in the sample plane. \textbf{C} View from the top: The orientation angle of the Janus particle that is relative to the heat source in the center is defined as $\gamma$.}
    \label{fig:sample_sketch}
\end{figure*}

\section*{Results}
\subsection*{Experimental System}
Our experimental system comprises two components suspended in a thin liquid water film (Fig.~\ref{fig:sample_sketch}B). The first component is a heat source realized by an immobilized silica particle (SiO$_2$) with radius $R_\mathrm{hs} = 1.3\,\si{\micro\metre}$ containing iron oxide particles (Fe$_3$O$_4$). The second component is a Janus particle (JP) consisting of a polystyrene sphere (radius $R_\mathrm{JP} = 1.15\,\si{\micro\metre}$) half-coated with a $50\,\si{\nano\metre}$ gold film. Both components are heated by separate $532\,\si{\nano\metre}$ lasers: the Janus particles via wide field illumination ($I =20-50\, \Iunit$) and the iron oxide particles via a focused beam controlled by a spatial light modulator ($I \approx 1000\,\Iunit$) (Fig.~\ref{fig:sample_sketch}A). This independent control allows a precise tuning of the relative heating to disentangle the various contributing physical effects that is hardly possible in other systems.

\begin{figure*}[ht]
    \centering
    \includegraphics[width=0.8\textwidth]{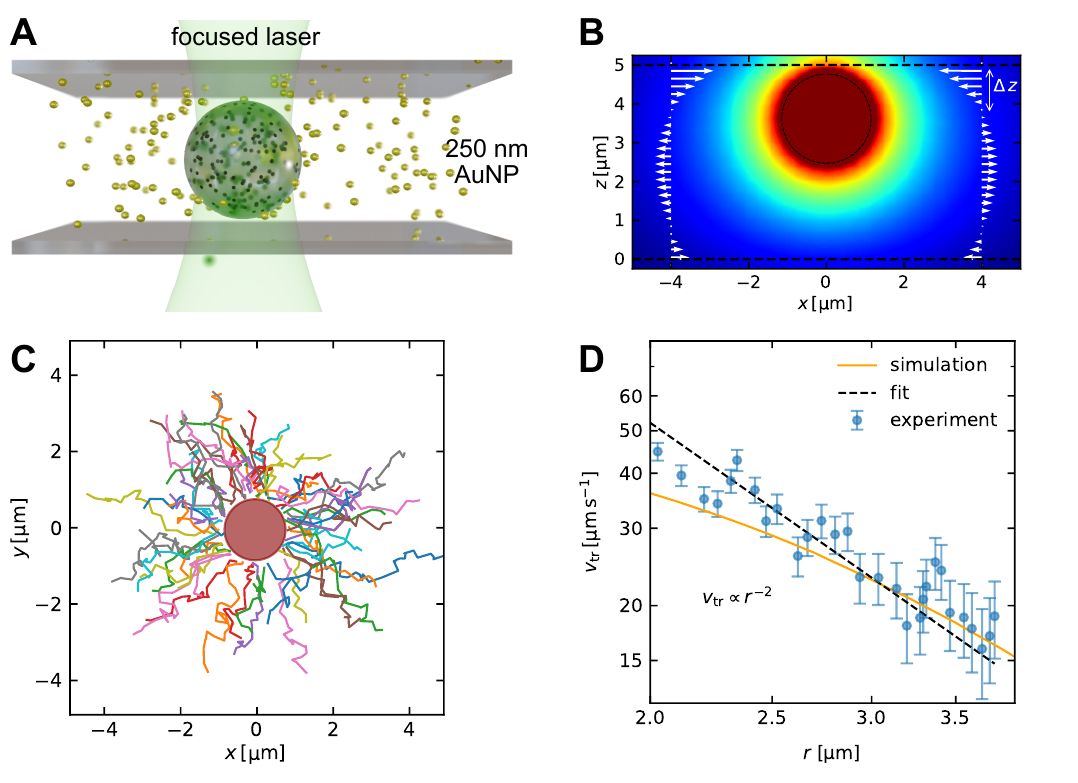}
    \caption{\textbf{Thermo-osmotic flow field of heat source} \textbf{A} Thermo-osmotic flow field experiment: Gold nanoparticles are added to the sample and are used to trace the flow field that is generated by the heat source. \textbf{B} Simulation of the thermo-osmotic flow field at the boundaries generated by the heated iron oxide particle shows the parabolic flow profile. $\Delta z$ indicates the distance over which the flow field reverses sign. \textbf{C} Trajectories of gold nanoparticle tracers reporting the thermo-osmotic flow towards the heat source at the substrate surface. The heat source is shown in red. \textbf{D} The radial velocity $v_\mathrm{tr}$ averaged over all trajectories as a function of radial distance from the heat source decays with $\propto r^{-2}$ as indicated by the fitted black dashed line. The error bars show the standard error of the radial velocity. The orange line shows that the result for the simulated flow field follows the same trend underlining the thermo-osmotic character of the flow field.}
    \label{fig:to_flows}
\end{figure*}

\subsection*{Heated Iron Oxide-Silica Particle}
The iron oxide particle is heated by a focused laser beam (Fig. \ref{fig:sample_sketch}B). It acts as a heat source (hs) in the sample and generates a temperature field that can lead to the thermophoresis of objects suspended in the liquid (e.g. Janus particles) and thermo-osmotic flows of the liquid induced at the container walls \cite{thermoosmotic_flows_bregulla, thermotaxisBregulla}.
To measure the thermo-osmotic flow field at the wall we add AuNPs of radius $a = 125\, \si{\nano\metre}$ to the sample solution (see Fig. \ref{fig:to_flows}A for a sketch) and track their trajectories. The gold nanoparticles are isothermal due to their high thermal conductivity and therefore do not exhibit thermophoretic drifts and only respond to the thermo-osmotic flow field. The measured trajectories in Fig.~\ref{fig:to_flows}C show that the tracers are attracted to the heat source in the plane of the upper cover slide. Analyzing the trajectories yields the radial velocity component $v_\mathrm{tr}$ close to the upper boundary as displayed in Fig. \ref{fig:to_flows}D as a function of the radial distance $r$ from the heat source. The experimental velocities follow the power law $v_\mathrm{tr}\propto r^{-2}$ as expected for the thermo-osmotic flow field at larger distances from the heat source.
The flow velocity of the thermo-osmotic flow field can be described by the Stokes equation using the quasi boundary slip velocity $\mathbf{u_s}$:

\begin{equation}
    \mathbf{u_s} = \mu_\mathrm{TO} \nabla T_\mathrm{hs} \approx - \mu_\mathrm{TO}\Delta T_\mathrm{hs} \frac{R_\mathrm{hs}}{r^{2}}\mathbf{\hat{e}_{r}}\,.
    \label{eq:tovelocit}
\end{equation}

Here $\nabla T_\mathrm{hs}$ denotes the temperature gradient generated at the glass/water boundary by the heated iron oxide particle. It can be approximated in the far field by the inverse square distance dependence of the temperature gradient of a point heat source, i.e. $\nabla T_\mathrm{hs} \approx - \Delta T_\mathrm{hs} R_\mathrm{hs}r^{-2}\mathbf{\hat{e}_{r}}$. The temperature of the heat source is experimentally obtained  by exploiting the phase transition of a liquid crystal (see Supplementary Note 1). The mobility coefficient $\mu_\mathrm{TO}$ in eq. \ref{eq:tovelocit} quantifies the strength of the thermo-osmotic flow field and is governed by the excess interactions at the liquid-solid interface as compared to the bulk liquid \cite{thermoosmotic_flows_bregulla,franzl2022hydrodynamic}. By fitting the experimental data in Fig. \ref{fig:to_flows}D to the form $v_\mathrm{tr} = m\,r^{-2} $ with m being the slope as indicated by the black dashed line, we can determine the thermo-osmotic mobility coefficient yielding a typical value of $\mu_\mathrm{TO} = m/(\Delta T_\mathrm{hs} \,R_\mathrm{hs}) =  1.54\, \si{\Muunit}$, which agrees well with the order of magnitude of previously found values for a water-glass interface $\mu_\mathrm{glass,water} = 0.6 - 4.4 $\, \si{\Muunit} \cite{thermoosmotic_flows_bregulla}. This mobility coefficient was also employed to simulate the flow field using finite element simulations (COMSOL), as illustrated in Fig. \ref{fig:to_flows}B. The orange curve shows the drift velocities of the tracer particles obtained by averaging the simulated flow field velocities over the distance in which the nanoparticles would diffuse quantified by the diffusion length $l_D = \sqrt{2 D_\mathrm{AuNP} \tau} = 440\,\si{\nano \metre}$. The translational diffusion coefficient of the gold nanoparticles is calculated as $D_\mathrm{AuNP} = k_\mathrm{B} T / 6\pi \eta a = 2.9 \, \si{\Dunit}$, where $T = 295\,\si{\kelvin}$ and $\eta = 1 \,\si{\milli\pascal\second}$. As illustrated in Fig. \ref{fig:to_flows}D, the calculated velocities closely mirror experimental findings (see Supplementary Note 3.1 for simulation methodology), reinforcing confidence in the derived coefficient.
\\
\begin{figure*}
    \centering
    \includegraphics[width=0.95\textwidth]{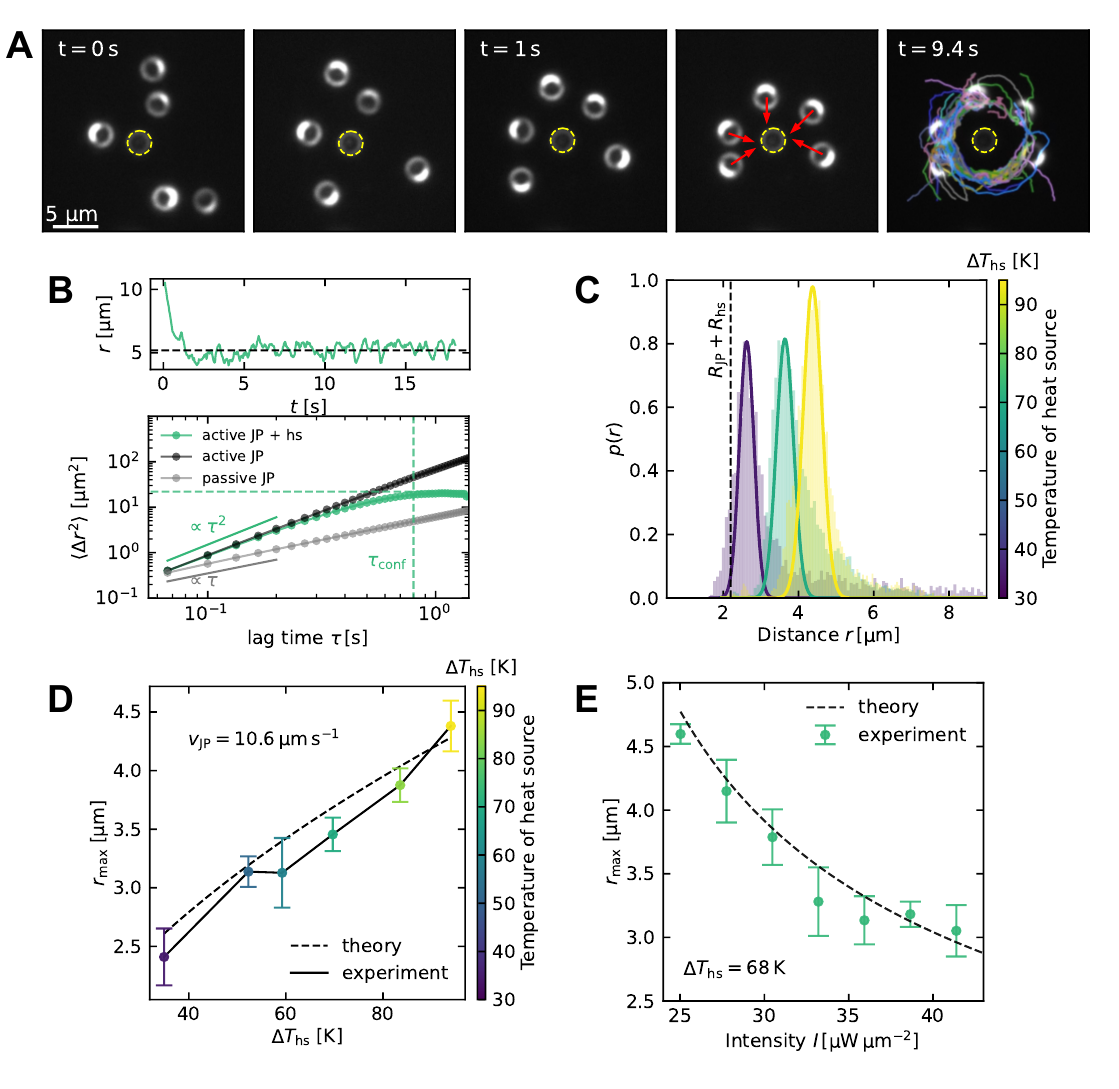}
    \caption{\textbf{Spatial Confinement of active Janus particles} \textbf{A} Dark field images in a time sequence: Janus particles self-organize into a structure around a heated particle in the center (yellow dashed circle) for a time $t> \tau_\mathrm{rot} = 9.38\, s$. \textbf{B} Top: The distance of a JP as a function of time shows that the JP keeps a constant distance. Bottom: MSD for passive JPs and active JPs with and without a heat source in the vicinity. The green data points show that active JPs are confined around the heat source after $\tau_\mathrm{conf} = 0.8\,\si{\second}$. \textbf{C} Probability distribution of the distances $r$ between JPs and heat source for different temperatures of the heat source. The lines correspond to the fit of the analytical density profile $\rho(r)$ with fitting parameters $\mu_\mathrm{TP,fit} = 3.11\,\si{\Muunit} $ and $\mu_\mathrm{TO,fit} = 1.51\,\si{\Muunit} $. \textbf{D} With increasing $\Delta T_{\mathrm{hs}}$ the distance $r_\mathrm{max}$ of the Janus particles to the heat source increases. The swimmers' velocity was kept constant to $ v_{\mathrm{JP}} = 10.6 \, \si{\Vunit}$.  \textbf{E} For a constant $\Delta T_{\mathrm{hs}} = 68\,\mathrm{K}$, the distance $r_\mathrm{max}$ decreases with the laser light intensity of the wide field illumination and thus with increasing propulsion speed $v_{\mathrm{JP}}$.}
    \label{fig:distance_dep}
\end{figure*}

\subsection*{Spatial confinement of Janus Particles}
The flow field generated at the boundary is directed towards the heat source yielding an attractive force on the Janus particles in the vicinity of the heat source as indicated by eq. \ref{eq:tovelocit}. For passive Janus particles we find an attraction of the JP to the heat source exhibiting a rolling motion when they are in close contact and under strong confinement (see Supplementary Note 3.2). This provides direct evidence for the presence of a significant thermo-osmotic flow field at the substrate, to which the JPs are exposed to. Our finding is in contrast to previous reports where passive JPs were repelled from a gold nanoparticle heat source due to the anisotropic thermophoresis of the JPs \cite{thermotaxisBregulla}. The thermophoresis of the JP is governed by an interfacial temperature gradient along the particle liquid interface $\nabla T_\mathrm{hs}$ created by the heat source and the thermo-osmotic mobility coefficient $\mu_\mathrm{TP}(\theta,\phi)$ which carries the asymmetry of the surface properties. While our JPs have equivalent surface mobility coefficients, the difference in size of the heat source is suggested to create a weaker contributions of the thermo-osmotic flow field in the previous study.\\
The behavior of the JPs undergoes a significant change upon activation of their propulsion mechanism by increasing the intensity of the wide field illumination. In this active state, the Janus particles begin to orbit around the heat source. Figure \ref{fig:distance_dep}A illustrates this dynamic behavior through a time-series of dark field microscopy images. In these images, the Janus particles are easily distinguishable by their strong scattering gold caps, while the central heat source, being less intense, is indicated by a yellow dashed circle for clarity. Figure \ref{fig:distance_dep}B (top graph) shows an example of the distance of one JP over time approaching the heat source from $r =10\, \si{\micro\metre}$ distance then staying at about $5\, \si{\micro\metre}$ distance for time periods considerably longer than the rotational diffusion time $\tau_\mathrm{rot} =9.38\,\si{\second}$. The rotational diffusion time $\tau_\mathrm{rot}$ is calculated using the equation $\tau_\mathrm{rot} =1/D_\mathrm{R}=8\pi \eta R_\mathrm{JP}^3 / k_\mathrm{B} T$, where $T = 295\,\si{\kelvin}$, $\eta = 1 \,\si{\milli\pascal\second}$, and $R_\mathrm{JP} = 1.15\, \si{\micro\metre}$. The confinement of the JPs' trajectories is further evidenced by the mean squared displacement (MSD) shown in the lower panel of Fig. \ref{fig:distance_dep}B. The MSD exhibits quasi-ballistic behavior at short times ($\propto \tau^{2}$) but reaches a plateau at longer lag times $\tau > \tau_\mathrm{conf} = 0.8\,\si{\second}$ when the JPs are propelled with a laser intensity of $ I = 40\, \si{\Iunit}$ and the central iron oxide particle is heated. This plateau quantifies the spatial confinement of the attracted Janus particles with $\sqrt{\langle \Delta r^2 \rangle (\tau >\tau_\mathrm{conf})} = 4.5 \,\si{\micro\metre}$. In contrast, passive JPs exhibit typical Brownian motion, characterized by a linear increase in the MSD over time. By fitting the MSD of the active JPs to $4D\tau + v_\mathrm{JP}^2 \tau^2$ for the ballistic regime with the translational diffusion coefficient $D = k_\mathrm{B} T / 6\pi \eta R_\mathrm{JP} = 0.20 \,\si{\micro \metre^2 \second^{-1}}$, where $T = 295\,\si{\kelvin}$, $\eta = 1 \,\si{\milli\pascal\second}$, and $R_\mathrm{JP} = 1.15\, \si{\micro\metre}$, the propulsion speed of the Janus particles was determined to be $v_\mathrm{JP} = 10.6\, \si{\Vunit}$. \\
As the Janus particles orbit the heat source, they maintain a characteristic distance $r$. Figure \ref{fig:distance_dep}C illustrates the probability density distributions of these distances for three distinct heating powers of the focused laser beam, corresponding to three different temperatures $\Delta T_\mathrm{hs}$ of the heat source. Throughout these measurements, the wide field illumination intensity was maintained at $ I = 40\, \si{\Iunit}$, ensuring a constant swimming speed of $ v_{\mathrm{JP}} = 10.6 \, \si{\Vunit}$ for the JPs. As $\Delta T_\mathrm{hs}$ increases, the maxima of these distributions shift towards larger distances. This relationship is quantitatively analyzed in Fig. \ref{fig:distance_dep}D, where the locations of the distribution maxima $r_\mathrm{max}$ are plotted over $\Delta T_\mathrm{hs}$. Keeping the temperature of the heat source constant at $\Delta T _\mathrm{hs} = 68\,\si{\kelvin}$ while increasing the wide field light intensity for the JPs propulsion yields a decreasing distance of the JPs to the heat source (Fig. \ref{fig:distance_dep}E).\\
The results presented thus far reveal three key observations:
i) Active JPs are attracted to the heat source, in contrast to passive particles which are repelled when only thermo-phoresis is active.
ii) The steady-state orbital distance of the active JPs is not simply the sum of the particle radii, but rather a dynamic parameter that varies with both the temperature of the heat source and the propulsion velocity of the Janus particles.
iii) Active JPs circle the heat source for periods significantly exceeding their rotational diffusion time. This extended orbital behavior indicates a sustained polarization towards the heat source when propelled, contrasting sharply with the rolling motion observed for passive JPs.
These findings collectively suggest a complex interplay of forces: their intrinsic activity, thermo-osmotic attraction and thermo-phoretic repulsion. This balance of non-equilbrium interactions confines the swimmers at a characteristic distance from the heat source (cf. Fig. \ref{fig:distance_dep}) facilitating their self-organization around the heated iron oxide particle. The total Janus particle flux density $\mathbf{j}_\mathrm{total}$ in the system incorporates the thermo-osmotic, thermo-phoretic, self-propulsion, and the diffusive contributions:
\begin{align}
\label{eq:total_flux}
    \mathbf{j}_\mathrm{total} &= \mathbf{j}_\mathrm{TP} + \mathbf{j}_\mathrm{TO} + \mathbf{j}_\mathrm{propulsion} + \mathbf{j}_\mathrm{D}\\
     & = -\rho(r) \mu_\mathrm{TP} \nabla T_\mathrm{hs} + \rho(r) \mu_\mathrm{TO} \nabla T_\mathrm{hs}  +\rho(r) \mathbf{v}_\mathrm{JP}  - D\nabla \rho(r) \nonumber \,.
\end{align}
Here, $\mu_\text{TP}$ and $\mu_\text{TO}$ are the thermo-phoretic and thermo-osmotic mobility coefficients respectively, $\rho(r)$ the JP number density profile, $r$ is the distance from the JP's center to the heat source's center and $\mathbf{v}_\mathrm{JP}$ the JP's velocity vector. We assume the JPs experience the undisturbed temperature gradient $\nabla T_\mathrm{hs}$ generated by the heat source, evaluated at the JP's geometric center (see Supplementary Note 4 for a detailed assessment). Further, temperature gradients causing thermo-phoretic drifts and osmotic flows differ only slightly at larger distance from the heat source. At steady state, spatial confinement of JP trajectories implies $\mathbf{j}_\mathrm{total} = 0$. Solving the differential equation in eq. \ref{eq:total_flux} for $\rho(r)$ gives the radial density profile:
\begin{align}
    \rho(r) = \rho_0 \exp\left( -\frac{1}{D} \left(  (\mu_\mathrm{TP} - \mu_\mathrm{TO}) \frac{\Delta T_\mathrm{hs}R_\mathrm{hs}}{r} -v_\mathrm{r} \,r\right) \right)\,,
\end{align}
with $v_\mathrm{r}$ being the radial component of the JP's velocity vector $\mathbf{v}_\mathrm{JP}$ relative to the heat source. The width of the distributions is determined by the translational diffusion coefficient $D = 0.20 \,\si{\micro \metre^2 \second^{-1}}$. The solid lines in Fig. \ref{fig:distance_dep}C show a fit of the analytical density profiles $\rho(r)$ by optimizing the parameters for the mobilities $\mu_\mathrm{TP}$ and $\mu_\mathrm{TO}$.  We find the best estimate for the phoretic mobility as $\mu_\mathrm{TP,fit} = 3.11\,\si{\Muunit}$ which is consistent with previous studies for JPs close to a heat source reporting a mobility coefficient in the same order of magnitude \cite{thermotaxisBregulla}. The thermo-osmotic mobility obtained from our model fit ($\mu_\mathrm{TO,fit} = 1.51\,\si{\Muunit}$) closely matches the value measured directly from thermo-osmotic flow field experiments ($\mu_\mathrm{TO} = 1.54\,\si{\Muunit}$). This agreement, along with the excellent match between our theoretical predictions and the measured distance distributions ($p(r)$) in Figure \ref{fig:distance_dep} C, validates that our model successfully describes the motion of Janus particles in the vicinity of the heat source.

The analytical solution for the location of the maximum of the distribution $r_\mathrm{max}$ to the heat source's center yields:
\begin{align}
    r_\mathrm{max} &  = \sqrt{\frac{ (\mu_\text{TP} - \mu_\text{TO} )\Delta T_\mathrm{hs} R_\mathrm{hs}}{ -v_\text{r} }}\, ,
    \label{analyticalR}
\end{align}
where $\mu_\mathrm{TP} > \mu_\mathrm{TO}$ and $v_\mathrm{r} < 0$. Detailed information on this model is given in the Supplementary Note 5. Assuming that the JPs point with the polystyrene side towards the heat source, we calculate $v_\mathrm{r} = - v_\mathrm{JP} = - 10.6\,\si{\Vunit}$. Using $\mu_\text{TP} = \mu_\text{TP,fit} = 3.11 \,\si{\Muunit}$ and our experimentally determined $\mu_\text{TO}$, we compare the analytical solution for $r_\mathrm{max}$ (eq. \ref{analyticalR}) with the measured distances in Fig. \ref{fig:distance_dep}D and \ref{fig:distance_dep}E. The black dashed lines represent the analytical solution. The close agreement between analytical and experimental results confirms that the spatial confinement of JPs arises from the interplay of attractive thermo-osmosis, repulsive thermo-phoresis, and the swimmer's self-propulsion.

\subsection*{Angular confinement of Janus particles}
\begin{figure*}
    \centering
    \includegraphics[width=1\textwidth]{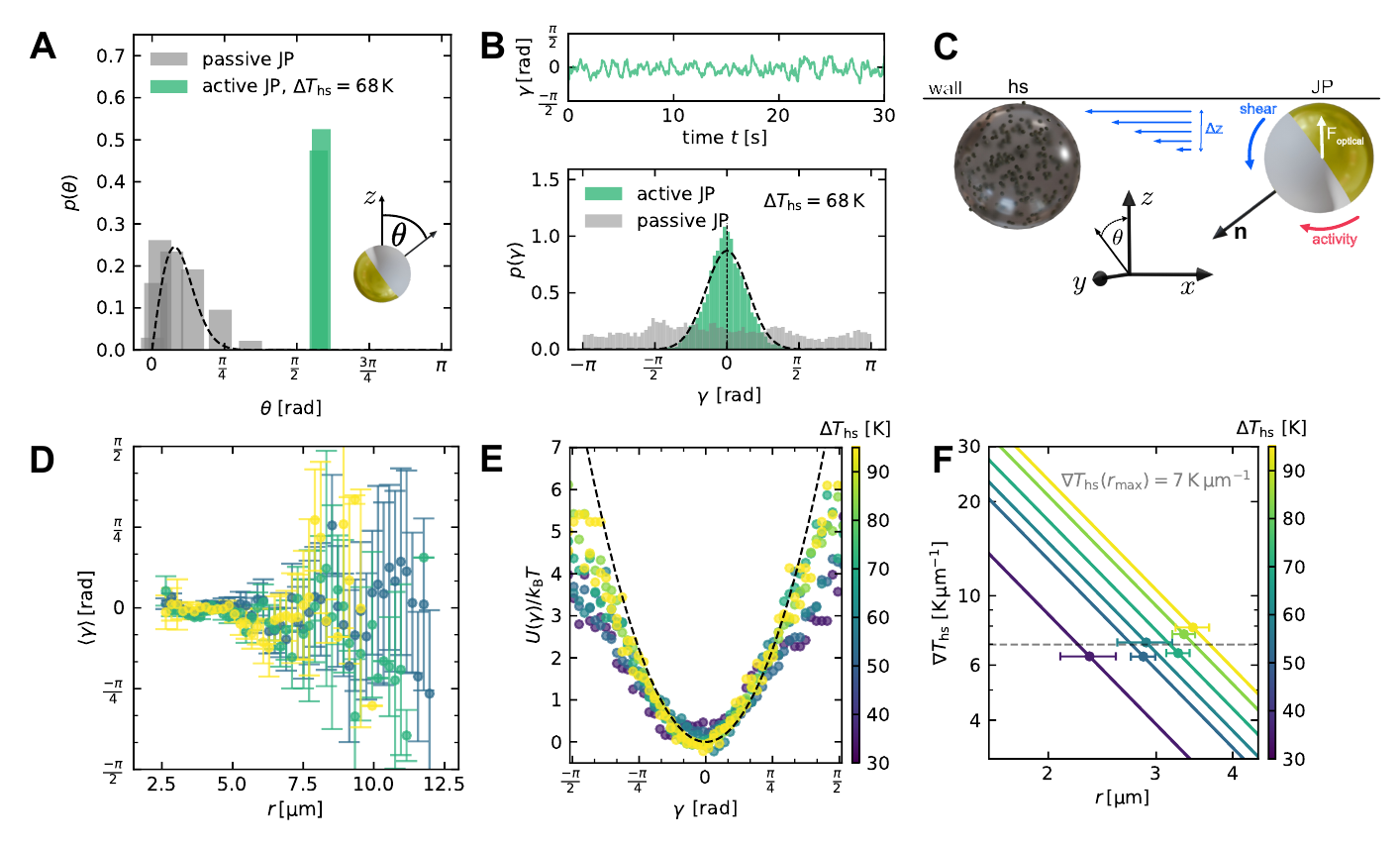}
    \makeatletter\long\def\@ifdim#1#2#3{#2}\makeatother
    \caption{\textbf{Angular confinement and polarization of Janus particles} \textbf{A} The probability density for the out-of-plane angle $\theta$ for the active JP around the heat source (green data) shows a strong alignment in the xz-plane as compared to passive Janus particles (grey data). \textbf{B} Top: $\gamma$ in the course of time points out that the JPs stay orientated towards the heat source for times longer than $\tau_\mathrm{rot} = 9.81\,\mathrm{s}$. Bottom: Probability distribution of the angle $\gamma$ for confined JPs (green data) shows higher probability that the JPs are pointing towards the heat source with $\gamma = 0$. The black dashed line shows a Gaussian fit with a zero mean and a variance of 0.21. For a passive Janus particle (grey data) the angles are equally distributed since there is no preferred orientation. \textbf{C} The thermo-osmotic flow field points into the  negative x-direction and creates a negative shear rate $\dot{\sigma} < 0$. \textbf{D} For distances $r > 7\,\mu \mathrm{m}$, the uncertainty of $\langle \gamma \rangle$ increases characterizing more random angles if the Janus particles are far away from heat source. \textbf{E} For all temperatures $\Delta T_\mathrm{hs}$ of the heat source, the potential of the relative angle $\gamma$ has the minimum at $\gamma = 0$ showing an in-plane polarization of the JPs. The black dashed line shows the analytical result for the effective potential. \textbf{F} The temperature gradient $\nabla T_\mathrm{hs}$ is shown as a function of radial distance $r$. The dots correspond to the distance $r_\mathrm{max}$ of the JP to the heat source. It shows that they are confined at the same temperature gradient $\nabla T_\mathrm{hs} (r = r_\mathrm{max}) = 7\,\si{\kelvin \per \micro \meter}$.}
    \label{fig:polarisation_dep}
\end{figure*}

Under the spatial confinement, the Janus particles orbit the heat source for times longer than the rotational diffusion time which suggest a polarized orientation towards the heat source. We analyzed the full 3D orientation of the Janus particles by considering the in-plane angle $\gamma$ relative to heat source's center and the out-of-plane angle $\theta$. The angle $\theta$ is defined relative to the positive z-axis, as illustrated in Fig. \ref{fig:polarisation_dep}A, and is determined by evaluating the overall intensity of the Janus particle throughout the experiment (see Supplementary Note 2) \cite{anthony2006single, anchutkin2024run}. Figure \ref{fig:polarisation_dep}A presents the probability density of $\theta$ for both active (green) and passive (grey) Janus particles. Active JPs exhibit a strong alignment in the xz-plane, moving along the upper boundary with a characteristic angle of $\theta = 1.8\,\si{rad}$. In contrast, passive JPs, being bottom-heavy, maintain a mean orientation of $\theta = 0\,\si{rad}$, indicating that their gold cap points downwards. The expected distribution $p(\theta)$ for passive JPs (dashed line in Fig. \ref{fig:polarisation_dep}A) is modeled using the gravitational potential $U_\mathrm{g} = -m g h_\mathrm{cm}\cos(\theta)$, with $m$ as the particles mass, $g$ as the gravitational acceleration and $h_\mathrm{cm}$ being the distance from the center of geometry to the center of mass with $h_\mathrm{cm} = R_\mathrm{JP} - R_\mathrm{cm} = 0.4\,\si{\micro \metre}$ for our Janus particles.

The in-plane angle $\gamma$, measured relative to the radial direction towards the heat source, exhibits a strong polarization for the active JPs close to the heat source. As shown in Fig. \ref{fig:polarisation_dep}B, these active JPs tend to align with $\gamma = 0$, meaning their polystyrene side preferentially faces the heat source. In contrast, passive particles, which roll along with the thermo-osmotic flow near the heat source, show no preferred orientation with $\gamma$ distributed uniformly across all angles. The active JPs' directional alignment becomes weaker when they are not in the steady state position at farther distance from the heat source, as shown by the increasing error bars in Fig. \ref{fig:polarisation_dep}D indicating the standard deviation of the angular distribution.
The distribution $p(\gamma)$ for active JPs is well approximated by a Gaussian (dashed line in Fig. \ref{fig:polarisation_dep}B, lower panel), suggesting an effective harmonic potential $U(\gamma)$ that confines the JPs' orientation. Figure \ref{fig:polarisation_dep}E illustrates this effective potential for various heat source temperatures, evaluated at the distances $r_\mathrm{max}$ (see Fig. \ref{fig:distance_dep}C). Notably, the potential's curvature shows hardly any dependence on the heat source temperature. Using the angular distribution $p(\gamma)$, the mean radial component of the swimming speed $\langle v_\mathrm{r} \rangle$ relative to the heat source is calculated by weighting the speed with the angular distribution according to $\langle v_\mathrm{r} \rangle = \int{\mathbf{v}_\mathrm{JP} \cdot \mathbf{\hat{e}}_\mathrm{r} p(\gamma) d\gamma} = - v_\mathrm{JP} \int{\cos(\gamma) \, p(\gamma) d\gamma} = - 0.9 \,v_\mathrm{JP}$ so that we can assume that $\langle v_\mathrm{r} \rangle \approx - v_\mathrm{JP} = - 10.6 \, \si{\Vunit}$. This validates the earlier assumption of $v_\mathrm{r} = - v_\mathrm{JP} = -10.6\,\si{\Vunit}$ used for calculating the analytical distances $r_\mathrm{max}$. \\
The observed angular confinement of the active Janus swimmers stems from a balance of multiple torques exerted on the particle. These torques result from different effects including hydrodynamic and activity induced interactions with the wall as well as an external radiation pressure as sketched in Fig. \ref{fig:polarisation_dep}C. The optical force originates from the laser light interacting with the JP's gold cap through scattering and absorption, generating a force parallel to the z-axis. This confines the JPs to the upper wall of the sample. Near the wall, two competing torques act on the JP around the y-axis: the radiation pressure tends to rotate the orientation vector away from the wall, while hydrodynamic interactions between the swimmer's self-generated flow field and the wall create a torque rotating the orientation vector towards the wall. The direction of this rotation aligns with predictions from lubrication theory, which describes particle rotation towards boundaries \cite{ruhle2018gravity, schaar2015detention, ishikawa2006hydrodynamic}. The Janus particle's self-generated flow field has a stagnation point at $2R_\mathrm{JP} = 2.3\,\si{\Vunit}$ from its geometric center, which is significantly further than the distance to the wall. Given this close proximity to the substrate, we cannot use the conventional far-field approximation that treats the active particle as a puller.\\
The thermo-osmotic flow field generated by the heat source also contributes to the angular confinement of the JPs. Due to its parabolic flow profile in the xz-plane which is shown in Fig. \ref{fig:to_flows}B, the JP will experience a shear with a negative shear rate $\dot{\sigma} < 0$. It will introduce a rotation of the orientation vector away from the wall around the y-axis. The stable out-of plane angle $\theta$ is therefore the balance of the torques exerted by the interaction with the wall, the radiation pressure and the shear of the hydrodynamic flow field as sketched in Fig. \ref{fig:polarisation_dep}C.\\

The Janus particles exhibit both strong confinement in their out-of-plane angle $\theta$ and alignment of their in-plane orientation $\gamma$ with the local hydrodynamic flow field. This alignment results in the particles swimming down the local flow field to the heat source, producing the polarization effect shown in Fig. \ref{fig:polarisation_dep}. When a Janus particle's in-plane angle $\gamma$ deviates from alignment with the heat source, it creates a perturbation $\Delta n_y$ in the orientation vector $\mathbf{n} = (n_x,n_y,n_z)$. According to the stability analysis in \cite{uspal2015rheotaxis}, this perturbation either decays or grows exponentially, depending on the signs of $n_x$ and $n_z$:
\begin{align}
    \Delta n_y = \Delta n_y (0) \exp{\frac{\dot{\sigma} f(h/R) n_z }{2 n_x} t}\,.
\end{align}
The stability of downstream swimming is now determined by the out-of-plane angle $\theta$, which sets the signs of $n_x$ and $n_z$ (see Supplementary Note 6 for details on the model). Our experimental analysis of the Janus particles' orientation shows that the JPs orient with their vectors pointing away from the wall ($n_z < 0$ and $n_x < 0$, see Fig. \ref{fig:polarisation_dep}C). Given that the shear rate $\dot{\sigma} < 0$, the perturbation $\Delta n_y$ decays exponentially with a characteristic relaxation time of $t_\mathrm{Rheo} = \frac{2 n_x}{\dot \sigma f n_z} = 1.29\,\si{\second}$. The shear rate is thereby calculated as $\dot{\sigma} = \langle v(r_\mathrm{max})\rangle_{\Delta z} / \Delta z = -10.6\,\si{\second^{-1}}$. Extracting the simulated flow velocities at $r=r_\mathrm{max}$ and averaging over the distance $\Delta z$, gives $\langle v(r_\mathrm{max})\rangle_{\Delta z} = - 13.98\,\si{\Vunit} $ whereas $\Delta z=1.31\, \si{\micro\metre}$ is the distance over which the thermo-osmotic flow field reverses sign (cf. Fig. \ref{fig:to_flows}B). Evaluating the temperature gradient of the heat source by $ \nabla T_\mathrm{hs}= \Delta T_\mathrm{hs} R_\mathrm{hs} / r^2$ at the  JPs stable positions $r = r_\mathrm{max}$ shows that the JPs all experience the same gradient of $\nabla T_\mathrm{hs} (r = r_\mathrm{max}) = 7\,\si{\kelvin \per \micro \meter}$ (cf. Fig. \ref{fig:polarisation_dep}F). Since $t_\mathrm{Rheo}$ is much shorter than the rotational diffusion time $\tau_\mathrm{rot} = 9.81\,\mathrm{s}$, hydrodynamic torques realign the JP much faster than rotational diffusion can disrupt the in-plane alignment. This leads to the observed stable downstream swimming with consistent in-plane polarization.\\
The corresponding effective potential confining the in-plane angle $\gamma$ in Fig. \ref{fig:polarisation_dep}D is therefore approximately  harmonic
\begin{align}
    U(\gamma)/k_\mathrm{B}T = \frac{1}{2} \kappa\, \gamma^2 = \frac{1}{2} \frac{1}{D_\mathrm{R}\, t_\mathrm{Rheo}} \gamma^2\,,
\end{align}
with $D_\mathrm{R}=(k_\mathrm{B} T)/8\pi \eta R_\mathrm{JP}^3 = 0.1 \,\si{\Drunit}$, where $T = 295\,\si{\kelvin}$, $\eta = 1 \,\si{\milli\pascal\second}$, and $R_\mathrm{JP} = 1.15\, \si{\micro\metre}$ following the derivations in \cite{uspal2015rheotaxis}. The black line in Fig. \ref{fig:polarisation_dep}D shows the obtained analytical result, which matches with the experimental data for the effective potential for $\gamma$ nicely. Overall this agreement unravels the interplay of the different non-equilibrium forces to confine position and to regulate the in-plane orientation of the active particle.

\section*{Discussion}
The polarization mechanism we observe represents a genuine example of self-regulation in active particles, emerging from a balance between two non-equilibrium fluxes that depend directly on the temperature gradient. This balance naturally guides the Janus particle to a location of constant, non-zero temperature gradient. The magnitude of this gradient is determined by the ratio of the active particle speed to the difference between the thermo-phoretic and thermo-osmotic mobilities: $\nabla T_\mathrm{hs} (r_\mathrm{max})=  v_\mathrm{JP} /(\mu_\mathrm{TP} - \mu_\mathrm{TO}) =  6.8 \,\si{\kelvin \,\micro \metre^{-1}}$ (cf. eq. \ref{analyticalR}). Our experimental measurements confirm this prediction, showing that particles consistently position themselves where they experience a temperature gradient of approximately $7\,\si{\kelvin \,\micro\metre^{-1}}$ when propelling at $ v_{\mathrm{JP}} = 10.6\, \si{\Vunit}$ (Fig. \ref{fig:polarisation_dep}F). This self-positioning ensures that particles experience a constant shear rate $\dot{\sigma}$ from the thermo-osmotic flow field, resulting in remarkable stability of the particle polarization across different heat source temperatures (Fig. \ref{fig:polarisation_dep}E). Thus, the particle's spatial position serves as an internal regulatory mechanism for maintaining its polarization state.

The robustness of this regulatory mechanism depends on two key conditions. First, the relaxation time $t_\mathrm{Rheo}$ must be shorter than the particle's rotational diffusion time, which restricts this effect to particles with sizes $R_\mathrm{JP}>0.6\, \si{\micro\metre}$. Second, the active particle's self-propulsion, which determines both the shear rate at the stable position and the polarization relaxation time, must be sufficient large for the studied Janus particle radius. To ensure the relaxation time remains shorter than the rotational diffusion time, the shear rate must exceed a critical value of $\dot{\sigma}_\mathrm{cr} = - 1.4  \,\si{\second^{-1}}$. This critical shear rate corresponds to a temperature gradient of $\nabla T_\mathrm{hs} = 1.2\,\si{\kelvin\per\micro\metre}$, requiring active particles to maintain a minimum speed of $v_\mathrm{JP}=2.3\, \Vunit$.

Regulatory mechanisms in active matter systems have traditionally been achieved through external control or complex feedback loops. These include sensing-based regulation \cite{franzl2020active,khadem2019delayed,alvarez2021reconfigurable}, communication-driven collective optimization \cite{zampetaki2021collective}, and externally controlled feedback systems \cite{bauerle2018self, khadka2018active}. While theoretical studies have explored the role of self-regulation in structure formation \cite{gopinath2012dynamical,baskaran2012self}, experimental demonstrations of internal regulation at the single-particle level have remained scarce.

Our work demonstrates how regulation can emerge purely from the interplay of physical forces, without external feedback or control. The precise experimental control over both the heat source and active particles allows us to isolate and understand the contributing mechanisms, particularly highlighting the crucial role of hydrodynamic interactions. While previous studies have shown how active particles respond to uniform flow fields through rheotaxis \cite{uspal2015rheotaxis, upstreamrheotaxis}, our system reveals more complex dynamics arising from the combination of hydrodynamic boundary interactions, thermo-osmotic effects, and thermophoresis. A key advantage of our thermally-driven approach is the precise controllability and measurability of the induced fields, contrasting with the challenges in controlling chemical gradients. While other active matter systems employ electro-osmotic and chemo-osmotic propulsion \cite{bazant2004induced, thome2023electrokinetic, moran2017phoretic, liebchen2021interactions}, individual contributions and the role of hydrodynamic interactions are difficult to disentangle.

This increased complexity present in our system, rather than being a limitation, enables new functionalities through local environmental control. Unlike studies using global flow fields \cite{si2020self, rubio2021self, baker2019fight}, our localized temperature control creates specific conditions for internal regulation mechanisms to emerge. This approach opens new possibilities for designing active matter systems with autonomous regulatory capabilities.

\section*{Materials and Methods}
The Janus particles (JP) are  $ 2.3\, \si{\micro\metre} $ diameter polystyrene beads (microParticles GmbH) that are capped with a $50\, \si{\nano \metre}$ thin gold hemisphere. As a heat source, a silica particle with a diameter of $2.54\,\si{\micro\metre} $ containing embedded iron oxide (SiO$_2$ Fe$_2$O$_3$) are used and added to the solution. The sample consists of a $ 5\, \si{\micro\metre}$ thin water film confined by two microscope cover slips that are coated with Pluronic F127 to prevent particle adsorption. The sample thickness was controlled by using polystyrene beads with a diameter of $ 5\,\si{\micro\metre}$. The sample is sealed with polydimethylsiloxane to avoid evaporation.  The particle motion was observed with an inverted microscope (Olympus IX71) under dark field illumination with an oil-immersion dark field condenser (Olympus, NA = 1.2) and an oil-immersion objective (Olympus UPlanApo ×100/0.6). It was recorded by a  sCMOS camera (Hamamatsu) with an exposure time of $33 \,\si{\milli\second}$. The dark field illumination provides an image of the Janus particles, that clearly shows the scattered light of the particles' gold cap. Using a custom-built tracking routine (TrackerLab \cite{trackerlab}), the Janus particles' x- and y-coordinates and also the in-plane angles $\phi$ are extracted. \\
For illuminating the heat source (SiO$_2$ Fe$_3$O$_2$ particle), a laser with a wavelength of $\lambda = 532\,\si{\nano \metre}$ and a beam waist of $0.6\, \si{\micro\metre}$ was focused into the sample plane (cf. Fig. \ref{fig:sample_sketch}A, focused illumination). Throughout the experiment, the power of the focused laser was varied from $I = 1100 \,\si{\Iunit}$ up to $I = 3100\,\si{\Iunit}$. For the thermophoretic self-propulsion of the Janus particles, a second laser with a wavelength of $\lambda = 532\,\si{\nano \metre}$ was incident from below providing a wide field illumination of the sample (cf. Fig.\ref{fig:sample_sketch}A, wide field illumination). The laser light intensity in the sample plane was varied between $20 - 50 \, \si{\Iunit}$ independent from the intensity of the focused laser. The schematic optical setup in Fig. \ref{fig:sample_sketch}A shows that the two laser paths are independent from each other and are coupled together right before entering the objective via a beam splitter.

\section*{Acknowledgement}
L.~R. and F.~C acknowledge financial support from the German Research Foundation (Deutsche Forschungsgemeinschaft, DFG) through project no 432421051.
The authors also acknowledge the careful proofreading of the manuscript by Andrea Kramer.

\section*{Author Contributions}
L.~R.~and F.~C. designed the experiments. L.~R.~carried out the experiments and simulations. L.~R.~and F.~C.~discussed the results and wrote the manuscript. D.~J.~Q. and D.~P. contributed to the measurements of the thermo-osmotic flow fields.

\section*{Competing interests}
The Authors declare no Competing Financial or Non-Financial Interests.


\end{document}